\documentclass[reprint, superscriptaddress, amsmath,amssymb,pra]{revtex4-1}

\usepackage{color}
\usepackage[colorlinks,urlcolor=blue,linkcolor=blue,anchorcolor=blue,citecolor=blue]{hyperref} 
\usepackage{float}
\usepackage{algorithm}
\usepackage{algpseudocode}
\usepackage{amsmath}
\usepackage{braket}
\usepackage{graphicx}
\usepackage{dcolumn}
\usepackage{bm}

\def\Loss{\operatorname{Loss}}
\def\dim{\operatorname{dim}}
\def\train{\operatorname{train}}
\def\poly{\operatorname{poly}}

\newcommand{\rhoop}{\hat{\rho}}
\newcommand{\Iop}{\hat{I}}
\newcommand{\fidelity}{\mathcal{F}}
\newcommand{\infidelity}{\mathcal{I}}
\newcommand{\trace}{{\rm tr}}

\makeatletter
\newenvironment{breakablealgorithm}
  {
   \begin{center}
     \refstepcounter{algorithm}
     \hrule height.8pt depth0pt \kern2pt
     \renewcommand{\caption}[2][\relax]{
       {\raggedright\textbf{\ALG@name~\thealgorithm} ##2\par}%
       \ifx\relax##1\relax 
         \addcontentsline{loa}{algorithm}{\protect\numberline{\thealgorithm}##2}%
       \else 
         \addcontentsline{loa}{algorithm}{\protect\numberline{\thealgorithm}##1}%
       \fi
       \kern2pt\hrule\kern2pt
     }
  }{
     \kern2pt\hrule\relax
   \end{center}
  }
\makeatother

\begin{document}

\preprint{APS/123-QED}

\title{Density matrix reconstruction using non-negative matrix product states}


\author{Donghong Han}

\affiliation{
 Institute of Fundamental and Frontier Sciences,\\
 University of Electronic Science and Technology of China, Chengdu, 610051, China
}%

\author{Chu Guo}
\email{guochu604b@gmail.com}
\affiliation{%
 Key Laboratory of Low-Dimensional Quantum Structures and Quantum Control of Ministry of Education,\\
 Department of Physics and Synergetic Innovation Center for Quantum Effects and Applications,\\
 Hunan Normal University, Changsha 410081, China\\
}%

\author{Xiaoting Wang}
\email{xiaoting@uestc.edu.cn}
\affiliation{
 Institute of Fundamental and Frontier Sciences,\\
 University of Electronic Science and Technology of China, Chengdu, 610051, China
}%

\date{\today}

\begin{abstract}

Quantum state tomography is a key technique for quantum information processing, but is challenging due to the exponential growth of its complexity with the system size. In this work, we propose an algorithm which iteratively finds the best non-negative matrix product state approximation based on a set of measurement outcomes whose size does not necessarily grow exponentially. Compared to the tomography method based on neural network states, our scheme utilizes a so-called tensor train representation that allows straightforward recovery of the unknown density matrix in the matrix product state form. As applications, the effectiveness of our algorithm is numerically demonstrated to reconstruct the ground state of the XXZ spin chain under depolarizing noise.

\end{abstract}

\maketitle

\section{\label{sec:introduction}introduction}

Characterizing an unknown quantum state is of central importance in developing quantum technologies. Standard quantum state tomography (QST) reconstructs a generic quantum state by performing projective measurements on an informationally complete basis~\cite{white_nonmaximally_1999, JamesWhite2001}. The number of projective measurements required grows exponentially with the system size. In the meanwhile, current quantum technologies have pushed the number of qubits to close to one hundred~\cite{arute_quantum_2019, WuPan2021, ZhuPan2022}, and scalable quantum state tomography schemes are in great need. With additional assumptions on the underlying quantum state, more efficient schemes than the standard QST have been proposed. For example, QST for a sparse quantum state by compressive sensing~\cite{gross_quantum_2010,liu_experimental_2012, Smith_continuous_2013, Riofrio2017}, QST for quantum states which are permutationally invariant [\onlinecite{toth_permutationally_2010}, \onlinecite{moroder_permutationally_2012}], and QST for quantum states which can be efficiently represented with a low-depth parametric quantum circuit[\onlinecite{liu_variational_2020}]. In particular, QST methods based on tensor network representation \cite{cramer_efficient_2010,lanyon_efficient_2017} as well as neural network ansatz~\cite{torlai_neural-network_2018, torlai_latent_2018, carrasquilla_reconstructing_2019, ahmed_quantum_2021,smith_efficient_2021} are promising to extend QST to a much larger scale, and both approaches have been demonstrated on several tens of qubits based on synthetic data.

For an unknown $L$-qubit pure state that can be well approximated by a matrix product state (MPS), it is proved that a set of $O(L)$ local $n$-body reduced density matrices suffices to reconstruct the unknown state, where $n$ is a constant and independent of $L$ if the underlying state has bounded entanglement. Thus, only $O(\poly(L))$ number of measurements are required~[\onlinecite{cramer_efficient_2010}, \onlinecite{lanyon_efficient_2017}]. It has also been shown that a similar approach can be applied to reconstructing an unknown mixed state, with an additional assumption of its \emph{invertibility}~[\onlinecite{baumgratz_scalable_2013}]. However, such methods based on local density matrices are not easy to implement in practice, since (1) exact tomography of a series of local density matrices may be already hard, and (2) we can only reconstruct an approximation of each local density matrix from tomography using a finite number of measurements, and the approximation errors could accumulate and affect the overall tomography performance of the entire state. Another method based on an MPS ansatz is proposed using an unsupervised machine learning algorithm, where only global measurement data on a randomly prepared basis are required~\cite{wang_scalable_2020}. Such method however only considers the reconstruction of pure states. Neural network state based algorithms constitute another important class of heuristic approaches for QST with excellent precision and scalability in practice. Specifically, neural network states are used to model the pure states~\cite{torlai_neural-network_2018,rocchetto_learning_2018} and the density matrices~\cite{torlai_latent_2018} as classical neural networks; alternatively, they are used to model the output probability distributions~\cite{carrasquilla_reconstructing_2019} as classical neural networks. In the latter approach, it will generally be exponentially hard to further reconstruct the state as a vector or a density matrix from the probability distribution. In the former approach, one could efficiently compute amplitudes based on the trained neural network state, but for other tasks such as computing expectation values one still needs to perform a sampling process based on the trained neural network state, which may not be as convenient or efficient, if the underlying quantum state could be well represented as an MPS. Another possible drawback of the QST methods based on neural network states is, a priori, it is not clear which neural network representation is suitable for a given unknown state. 

\begin{figure*}[htp]
    \centering
    \includegraphics[width=0.85\linewidth]{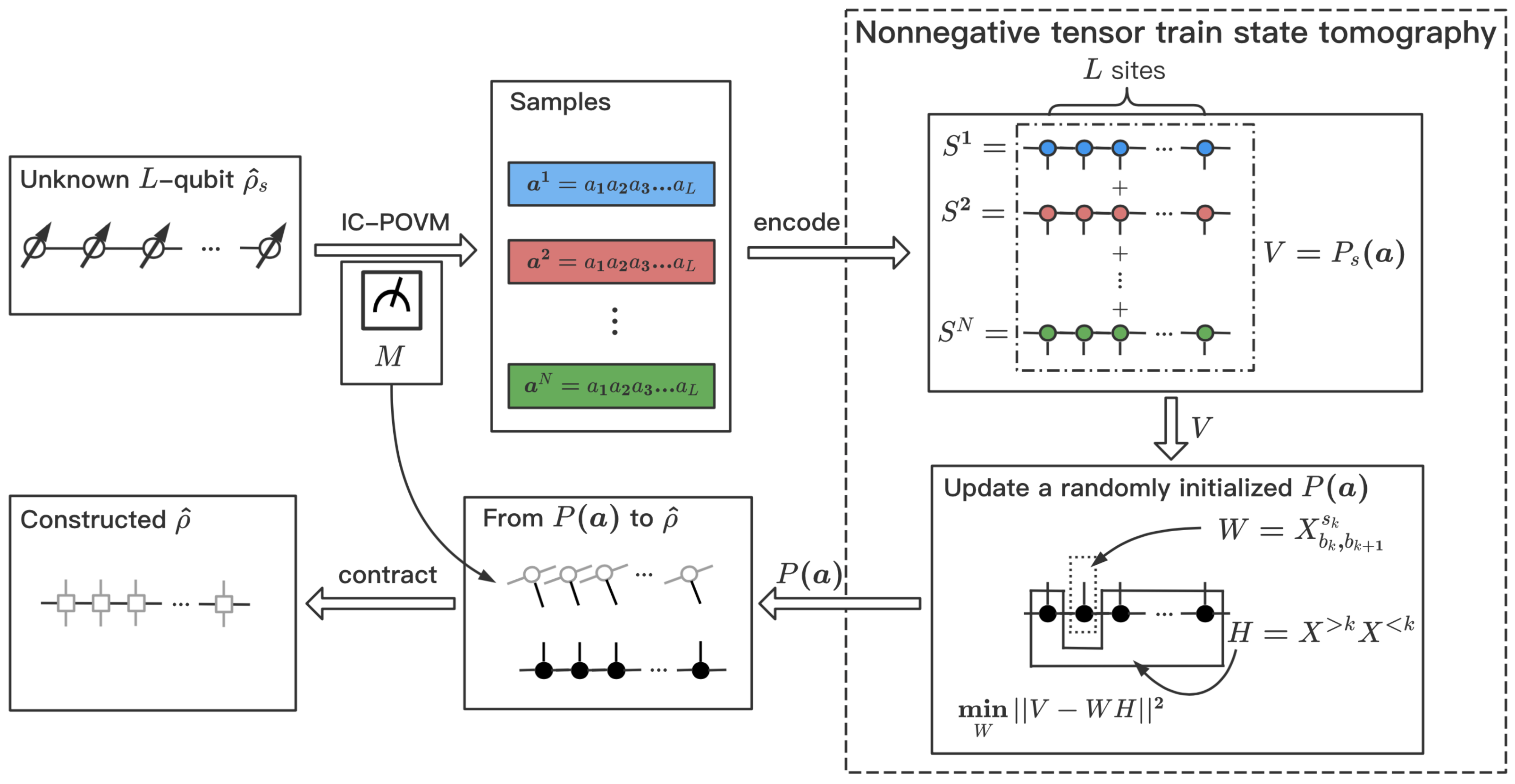}

\caption{A flowchart of our nonnegative tensor train state tomography algorithm. Given an unknown $L$-qubit quantum state $\rhoop_s$, an IC-POVM is performed to obtain a sample of bitstrings $\{\bm{a}^j\}$, each of which is then encoded into a \emph{one hot} MPS $S^j$. Based on these $S^j$, an optimal MPS $P(\bm{a})$ with a fixed bond dimension is found through optimization, satisfying that it is the closest to the superposition of all state $S^j$. Then an MPO $\rhoop$ is reconstructed, with the same bond dimension as $P(\bm{a})$, by applying the inverse of the IC-POVM locally on each site of $P(\bm{a})$. Simulations in this work are based on synthetic data.}

\label{fig:scheme}
\end{figure*}

Inspired by the fact that the MPS has been adapted to represent the multivariate probability distribution function, often referred to as the tensor train representation [\onlinecite{oseledets_tt-cross_2010}], we propose a QST scheme that combines the advantages of both the tensor network approach and the neural network approach. Specifically, in the first stage, a tensor train representation of the multivariate distribution function is constructed based on the quantum measurement data, instead of a neural network representation, and a density matrix renormalization group (DMRG)-like algorithm is used to find the optimal tensor train representation. After that, the tensor train is transformed back into a matrix product operator (MPO) representation of the unknown density matrix. Compared to the established QST methods based on tensor network states, our scheme directly uses a tensor train representation for the multivariate probability distribution function, instead of the unknown density matrix; compared to the established QST methods based on neural network states, our scheme constructs the unknown density matrix as an MPO, which usually allows more convenient and efficient evaluations of observables. The flowchart of our algorithm is summarized in Fig.~\ref{fig:scheme}. This work is organized as follows: we show the details of our QST scheme in Sec.~\ref{sec:method}, and then numerically demonstrate our algorithm for the ground state of the XXZ chain perturbed by depolarizing noise in Sec.~\ref{sec:results}, ended with a concluding discussion in Sec.~\ref{sec:conclusion}.


\section{\label{sec:method}Method}
For QST, we use quantum measurements given by informationally complete positive operator-valued measures (POVMs)~\cite{garcia-perez_learning_2021, Flammia_2006, Sosa_Experimental_2017}, which describes the most general quantum measurements allowed by quantum theory [\onlinecite{nielsen_quantum_2002}]. We denote the single qubit POVM as $\{M^s\}$, where $M^s$ is positive semi-definite satisfying $\sum_{s}M^s =\Iop$, where $\Iop$ is the identity matrix. For a single qubit, a minimal informationally complete POVM can be chosen as $M^s = \frac{1}{2}\ket{\psi^s}\bra{\psi^s}$ with
\begin{equation}
\label{equ:POVM}
	\begin{split}
\ket{\psi^0} &= \ket{0}, \\
\ket{\psi^1} &= \sqrt{\frac{1}{3}}\ket{0}+\sqrt{\frac{2}{3}}\ket{1}, \\
\ket{\psi^2} &= \sqrt{\frac{1}{3}}\ket{0}+\sqrt{\frac{2}{3}}e^{i\frac{2\pi}{3}}\ket{1}, \\
\ket{\psi^3} &= \sqrt{\frac{1}{3}}\ket{0}+\sqrt{\frac{2}{3}}e^{i\frac{4\pi}{3}}\ket{1},
\end{split}
\end{equation}
which form the vertices of a regular tetrahedron in the Bloch sphere [\onlinecite{renes_symmetric_2004}]. Such a single-qubit POVM, $\bm M=\{M^s\}_{s=0,1,2,3}$ can be viewed as a 3-dimensional tensor written as $M^{s}_{\sigma,\sigma'}$ with two physical indices $\sigma,\sigma'$ of dimension 2 and another index $s$ of dimension 4 corresponding to different measurement outcomes. If we reshape the single-qubit density matrix into a vector of size 4, then $\bm M$ becomes a $4\times4$ invertible matrix, representing a one-to-one mapping between the density matrix and the single-qubit probability distribution.

As in [\onlinecite{carrasquilla_reconstructing_2019}], for an $L$-qubit quantum system, we consider the quantum measurement defined by
\begin{equation}
    M^{\otimes \bm{a}}\equiv M^{a_1}\otimes M^{a_2} \otimes \dots \otimes M^{a_L} ,
\end{equation}
where $\bm{a}=(a_1,\ldots,a_L)$ represents a string of integers specifying the local projectors, and each integer $a_l \in \{0, 1, 2, 3\}$. The probability distribution $P(\bm{a})$ forms an $L$-variable distribution function in which each local dimension $d$ is equal to $4$, satisfying $P(\bm{a}) \ge 0$ and $\sum_{\bm{a}}P(\bm{a})=1$. Interestingly, if we assume that the unknown quantum state can be efficiently represented as an MPO
\begin{equation}
    \rhoop=\sum_{b_1,b_2,...,b_{L+1}}W^{\sigma_{1},\sigma_{1}'}_{b_1,b_2}W^{\sigma_2,\sigma_{2}'}_{b_2,b_3} \dots W^{\sigma_L,\sigma_{L}'}_{b_L,b_{L+1}} ,
\end{equation}
where $b_l$ denotes the auxiliary index, then $P(\bm{a})$ can be written as an MPS
\begin{equation}\label{equ:pa}
    P(\bm{a})=\sum_{b_1,b_2,...,b_{L+1}}X^{s_1}_{b_1,b_2}X^{s_2}_{b_2,b_3} \dots X^{s_L}_{b_L,b_{L+1}} ,
\end{equation}
with each tensor $X^{s_l}_{b_l,b_{l+1}}=\sum_{\sigma_l,\sigma_{l}'}W^{\sigma_l,\sigma_{l}'}_{b_l,b_{l+1}}M^{s_l}_{\sigma_l,\sigma_l '}$. Therefore the bond dimensions of $P(\bm{a})$, which are defined as the dimensions of the auxiliary indices $D_l = \dim(b_l)$, are exactly the same as the bond dimensions of $\hat{\rho}$. Since $\bm{M}$ is invertible, we can efficiently transform back and forth between the density matrix $\hat{\rho}$ and the probability distribution $P(\bm{a})$. Here we choose to first construct $P(\bm{a})$ as an MPS, and then transform it back into a density matrix. One advantage of this approach is that as long as the constructed $P(\bm{a})$ is a proper probability distribution, then the density matrix $\hat{\rho}$ from this approach will automatically be Hermitian (which however may not be positive if there are not enough measurement data).

Thus the problem reduces to reconstructing an approximate probability distribution $P_s(\bm{a})$, based on a set of $N$ samples $\bm{a}^1, \bm{a}^2, \dots , \bm{a}^N$ collected from experiment. Assuming among those samples there are only $N_s$ different ones, which are denoted as $\bm{a}^1, \bm{a}^2, \dots , \bm{a}^{N_s}$, where each distinct sample $\bm{a}^j$ has multiplicity $n_j$, and each $\bm{a}^j$ appears with a probability $P_s(\bm{a}^j)=\frac{n_j}{N}$, $j=1,\cdots,N_s$. $P_s(\bm{a}^j)$ will eventually converge to the exact distribution $P(\bm{a})$ as $N$ increases. For a limited number of samples, the entries of $P_s(\bm{a})$ are approximately equal to the corresponding entries of $P(\bm{a})$. Therefore the original QST is reduced to the following mathematical problem: given some approximated values of the nonzero elements of an unknown $P(\bm{a})$, how can we construct a tensor train approximation of $P(\bm{a})$ with a minimum bond dimension $D$?

To this end, we note that each sample $\bm{a}^j$ can be encoded as a \emph{one hot} MPS:
\begin{equation}\label{equ:mapatoA}
    \bm{a}^j \mapsto S^j = \sum_{c_1, c_2, \dots , c_{L+1}}A^{s_1}_{j,c_1,c_2}A^{s_2}_{j,c_2,c_3} \dots A^{s_L}_{j,c_L,c_{L+1}} ,
\end{equation}
such that $\dim(c_l)=1$ and $\dim(s_l)=4$ for all $1\leq l\leq L$, and each tensor $A^{s_l}_{j,c_l,c_{l+1}}$ satisfies $A^{s_l}_{j,0,0} = 1$ iff $s_l$ equals to $a_l$, and $0$ otherwise. For a specific $\bm{a}^j$, if $a_l$ in $\bm{a}^j$ is 3 then $A^{s_l}_{j,c_l,c_{l+1}}$ satisfies $A^{0}_{j,0,0} = 0$, $A^{1}_{j,0,0} = 0$, $A^{2}_{j,0,0} = 0$ and $A^{3}_{j,0,0} = 1$.
With Eq.~(\ref{equ:mapatoA}) we can rewrite the probability distribution formed by $N$ samples as
\begin{equation}\label{equ:psa}
    P_s(\bm{a}) = \sum^{N_s}_{j=1}\frac{n_j}{N}S^j,
\end{equation}
where $P_s(\bm{a})$ can be viewed as a superposition of all one hot states $S^j$, weighted by the multiplicities. $P_s(\bm{a})$ can be directly taken as the best approximation of $P(\bm{a})$, that is, setting $P(\bm{a})=P_s(\bm{a})$, $P_s(\bm{a})$ can be directly evaluated from Eq.~(\ref{equ:psa}) using simple MPS arithmetics. However, given a limited set of samples, the bond dimension of the resulting MPS could be much larger than that of the target distribution. Moreover, this might result in an over-fitting problem, since $P_s(\bm{a})$ will be perfect for known samples but will be $0$ for unknown samples. For better efficiency and generalizability, one can search for $P(\bm{a})$ which is approximately equal to $P_s(\bm{a})$ under the condition that the bond dimension is bounded by a fixed value $D$. This could be done in two approaches [\onlinecite{schollwock_density-matrix_2011}]: (1) evaluating Eq.~(\ref{equ:psa}) exactly and then compressing the resulting MPS using SVD, and (2) iteratively searching for the solution to the following optimization problem
\begin{equation}\label{equ:fpa}
    \Loss(P(\bm{a})) \equiv \sum^{N_s}_{j=1} \left \| P(\bm{a}) - P_s(\bm{a}) \right \|^2 ,
\end{equation}
with a maximal bond dimension $D$, where $ \|P \|$ denotes the Frobenius norm of the tensor $P$. We will follow the latter approach which is more precise in practice. One complication here is that if one directly uses the approaches in [\onlinecite{schollwock_density-matrix_2011}], the MPS ansatz will be kept in a canonical form by iteratively using either singular value decomposition (SVD) or QR decomposition, and the solution generally contains negative values, which is undesirable for a probability distribution. To ensure the nonnegativity of $P(\bm{a})$, one could represent $P(\bm{a})$ as a nonnegative MPS instead, that is, each tensor $X$ in Eq.~(\ref{equ:pa}) is non-negative. Several algorithms have been proposed to approximate a target probability distribution using a non-negative MPS with a fixed bond dimension [\onlinecite{lee_nonnegative_2016},\onlinecite{shcherbakova_nonnegative_2019},\onlinecite{shcherbakova_nonnegative_Large-Scale_2019}]. Here we use a refined approach based on [\onlinecite{shcherbakova_nonnegative_Large-Scale_2019}], the central idea of which is to use a non-negative matrix decomposition instead of SVD or QR decomposition.

Specifically, we first define the following tensors $X^{>k}_{b_{k+1};s_{k+1}, \dots ,s_L}$ and $X^{<k}_{s_{1}, \dots ,s_{k-1};b_k}$:
\begin{subequations}
\begin{align}
&X^{>k} = \sum_{b_{k+2}, \dots , b_{L+1}}X^{s_{k+1}}_{b_{k+1},b_{k+2}} \dots X^{s_{L}}_{b_{L},b_{L+1}},\label{equ:xgk}\\
&X^{<k} = \sum_{b_1, \dots , b_{k-1}}X^{s_{1}}_{b_{1},b_{2}} \dots X^{s_{k-1}}_{b_{k-1},b_{k}},\label{equ:xlk}
\end{align}
\end{subequations}
and $G^{>k}_{b_{k+1},b_{k+1} '}$ and $G^{<k}_{b_{k},b_{k} '}$:
\begin{subequations}
\begin{align}
    &G^{>k}=\sum_{s_{k+1}, \dots ,s_L} X^{>k}_{b_{k+1};s_{k+1}, \dots ,s_L}X^{>k}_{b_{k+1} ';s_{k+1}, \dots ,s_L}, \\
    &G^{<k}=\sum_{s_1, \dots ,s_{k-1}} X^{<k}_{s_1, \dots ,s_{k-1} ; b_k}X^{<k}_{s_1, \dots ,s_{k-1};b_k '}.
\end{align}
\end{subequations}
With Eq.(\ref{equ:xgk}) and (\ref{equ:xlk}) we can rewrite $P(\bm{a})$ as
\begin{equation}\label{equ:newpa}
    P(\bm{a}) = \sum_{b_k,b_{k+1}}X^{s_k}_{b_k,b_{k+1}} X^{<k}_{ \dots ,s_{k-1};b_k} X^{>k}_{b_{k+1};s_{k+1}, \dots },
\end{equation}
for each $1\leq k\leq L$. Substituting Eq.~(\ref{equ:newpa}) into Eq.~(\ref{equ:fpa}), the loss function becomes $\left\| V-WH \right\|^2$, with $V = P_s(\bm{a})$, $W = X^{s_k}_{b_k,b_{k+1}}$ and $H = X^{<k}_{ \dots ,s_{k-1};b_k}X^{>k}_{b_{k+1};s_{k+1}, \dots}$. Thus the goal is to find the best non-negative factorization of $V$. One of the most well known approaches for solving this problem is the following updating rule
\begin{subequations}
\begin{align}
    W\leftarrow W \circ \frac{[VH^t]}{[WHH^t]} , \label{equ:W}\\
    H\leftarrow H \circ \frac{[W^tV]}{[W^TWH]} , \label{equ:H}
\end{align}
\end{subequations}
where $\circ$ means Hadmard (element-wise) product and $\frac{[A]}{[B]}$ denotes the element-wise division of the matrices $A$ and $B$~\cite{lee_learning_1999}. With Eqs.(\ref{equ:W}) and (\ref{equ:H}) the loss function is guaranteed to decrease monotonically. From Eq.~(\ref{equ:W}), one can update the tensor $X^{s_k}_{b_k,b_{k+1}}$ as
\begin{equation}\label{equ:X}
    X^{s_k}_{b_k,b_{k+1}} \leftarrow X^{s_k}_{b_k,b_{k+1}} \circ \frac{[\sum_{s_{l\neq k}}P_s(\bm{a})X^{<k}X^{>k}]}{[\sum_{b_k',b_{k+1}'}X^{s_k}_{b_k',b_{k+1}'}G^{>k}_{b_{k+1},b_{k+1} '} G^{<k}_{b_{k},b_{k} '}]}
\end{equation}
The denominator on the right hand side of Eq.~(\ref{equ:X}) can be efficiently evaluated without computing the summation in Eq.~(\ref{equ:psa}), for which we define two tensors $\widetilde{G}^{>k}_{j,b_{k+1},c_{k+1}}$ and $\widetilde{G}^{<k}_{j,b_{k},c_{k}}$:
\begin{subequations}
\begin{align}
    &\widetilde{G}^{>k}_{j} = \sum_{s_{k+1}, \dots ,s_L} X^{>k}_{b_{k+1};s_{k+1}, \dots ,s_L}A^{>k}_{j,c_{k+1};s_{k+1}, \dots ,s_L}, \\
    &\widetilde{G}^{<k}_{j} = \sum_{s_1, \dots ,s_{k-1}} X^{<k}_{s_1, \dots ,s_{k-1}; b_k}A^{<k}_{j,s_1, \dots ,s_{k-1};c_k} ,
\end{align}
\end{subequations}
where $A^{>k}$ and $A^{<k}$ are defined similarly as $X^{>k}$ and $X^{<k}$ in Eq.~(\ref{equ:xgk}) and (\ref{equ:xlk}). Then we have
\begin{equation}
\begin{split}
    &\sum_{s_{l\neq k}}P_s(\bm{a})X^{<k}X^{>k} \\
    &=\sum_{j}\frac{n_j}{N}\sum_{b_k',b_{k+1}'}A^{s_k}_{j,c_k,c_{k+1}}\widetilde{G}^{>k}_{j,b_{k+1},c_{k+1}}\widetilde{G}^{<k}_{j,b_{k},c_{k}} .
\end{split}
\end{equation}
The complete algorithm to find the optimal $P(\bm{a})$ that minimizes $\Loss(P(\bm{a}))$ in Eq.~(\ref{equ:fpa}) is summarized in Algorithm~\ref{alg:Framwork}. Once $P(\bm{a})$ is found, the best MPO $\rhoop$ can be reconstructed by applying the inverse of the IC-POVM locally on each site of $P(\bm{a})$, as illustrated in Fig.~\ref{fig:scheme}. 

\begin{breakablealgorithm}
\caption{Nonnegative Tensor Train State Tomography}
\label{alg:Framwork}
  \begin{algorithmic}[1]
    \Require
      the set of samples from POVM measurement;
    \Ensure
      near-optimal non-negative MPS form of $P(\bm a)$;
    \State Encode each $\bm{a}^j$ into $A^{j}$ according to Eq.~(\ref{equ:mapatoA});
    \State Randomly initialize $P(\bm a)$ as in [\onlinecite{holtz_alternating_2012}];
    \label{code:RandomPa}
    \State $\widetilde{G}^{<1}_{j,b_{1},c_{1}}=1$,$\widetilde{G}^{>L}_{j,b_{L+1},c_{L+1}}=1$
    \State $G^{<1}_{b_{1},b_{1}'}=1$,$G^{>L}_{b_{L+1},b_{L+1}'}=1$

    \For{$k=1, 2, \dots ,L-1$}
        \State $G^{<k+1}_{b_{k+1},b_{k+1}'}=
        \sum_{s_k,b_k,b_k'} G^{<k}_{b_k,b_k'} X^{s_k}_{b_k,b_{k+1}} X^{s_k}_{b_k',b_{k+1}'} $;
        \For{$j = 1,2,\dots,N_s$}
            \State $\widetilde{G}^{<k+1}_{j,b_{k+1},c_{k+1}}= $
            \Statex \qquad \qquad \qquad \qquad $\sum_{s_k,b_k,c_k} \widetilde{G}^{<k}_{j,b_k,c_k} X^{s_k}_{b_k,b_{k+1}} A^{s_k}_{j,c_k,c_{k+1}} $;
        \EndFor
    \EndFor

    \For{$k=L-1,L-2,\dots, 1$}
        \State $G^{>k}_{b_{k+1},b_{k+1}'}= $
        \Statex \qquad $\sum_{s_{k+1},b_{k+2},b_{k+2}'}
        G^{>k+1}_{b_{k+2},b_{k+2}'} X^{s_{k+1}}_{b_{k+1},b_{k+2}} X^{s_{k+1}}_{b_{k+1}',b_{k+2}'} $;
        \For{$j = 1,2,\dots,N_s$}
            \State $\widetilde{G}^{>k}_{j,b_{k+1},c_{k+1}}= $
            \Statex \qquad $\sum_{s_{k+1},b_{k+2},c_{k+2}} \widetilde{G}^{>k+1}_{j,b_{k+2},c_{k+2}} X^{s_{k+1}}_{b_{k+1},b_{k+2}} A^{s_{k+1}}_{j,c_{k+1},c_{k+2}} $;
        \EndFor
    \EndFor

    \label{code:fram:GandGtilde}
    \While{$true$}
        \For{$k=1, 2, \dots ,L-1$}
            \State update $X^{s_k}_{b_k,b_{k+1}}$ using Eq.~(\ref{equ:X});
            \State $G^{<k+1}_{b_{k+1},b_{k+1}'}=
            \sum_{s_k,b_k,b_k'} G^{<k}_{b_k,b_k'} X^{s_k}_{b_k,b_{k+1}} X^{s_k}_{b_k',b_{k+1}'} $;
            \For{$j = 1,2,\dots,N_s$}
                \State $\widetilde{G}^{<k+1}_{j,b_{k+1},c_{k+1}}= $
                \Statex \qquad \qquad \qquad \qquad $\sum_{s_k,b_k,c_k} \widetilde{G}^{<k}_{j,b_k,c_k} X^{s_k}_{b_k,b_{k+1}} A^{s_k}_{j,c_k,c_{k+1}} $;
            \EndFor
        \EndFor
        
        \For{$k = L-1, L-2,\dots , 1$}
            \State update $X^{s_{k+1}}_{b_{k+1},b_{k+2}}$ using Eq.~(\ref{equ:X});
            \State $G^{>k}_{b_{k+1},b_{k+1}'}=$
            \Statex \qquad \quad $\sum_{s_{k+1},b_{k+2},b_{k+2}'} G^{>k+1}_{b_{k+2},b_{k+2}'} X^{s_{k+1}}_{b_{k+1},b_{k+2}} X^{s_{k+1}}_{b_{k+1}',b_{k+2}'} $;
            \For{$j = 1,2,\dots,N_s$}
                \State $\widetilde{G}^{>k}_{j,b_{k+1},c_{k+1}}= $
                \Statex \qquad \quad$\sum_{s_{k+1},b_{k+2},c_{k+2}} \widetilde{G}^{>k+1}_{j,b_{k+2},c_{k+2}} X^{s_{k+1}}_{b_{k+1},b_{k+2}} A^{s_{k+1}}_{j,c_{k+1},c_{k+2}} $;
            \EndFor
        \EndFor
        
        \If {\emph{stopping criterion is meet}}
            \State break;
        \EndIf
    \EndWhile
    \State \Return $P(\bm{a})$ \label{code:recentEnd}
  \end{algorithmic}
\end{breakablealgorithm}

\begin{figure}[htp]
    \centering
    \includegraphics[width=0.95\columnwidth]{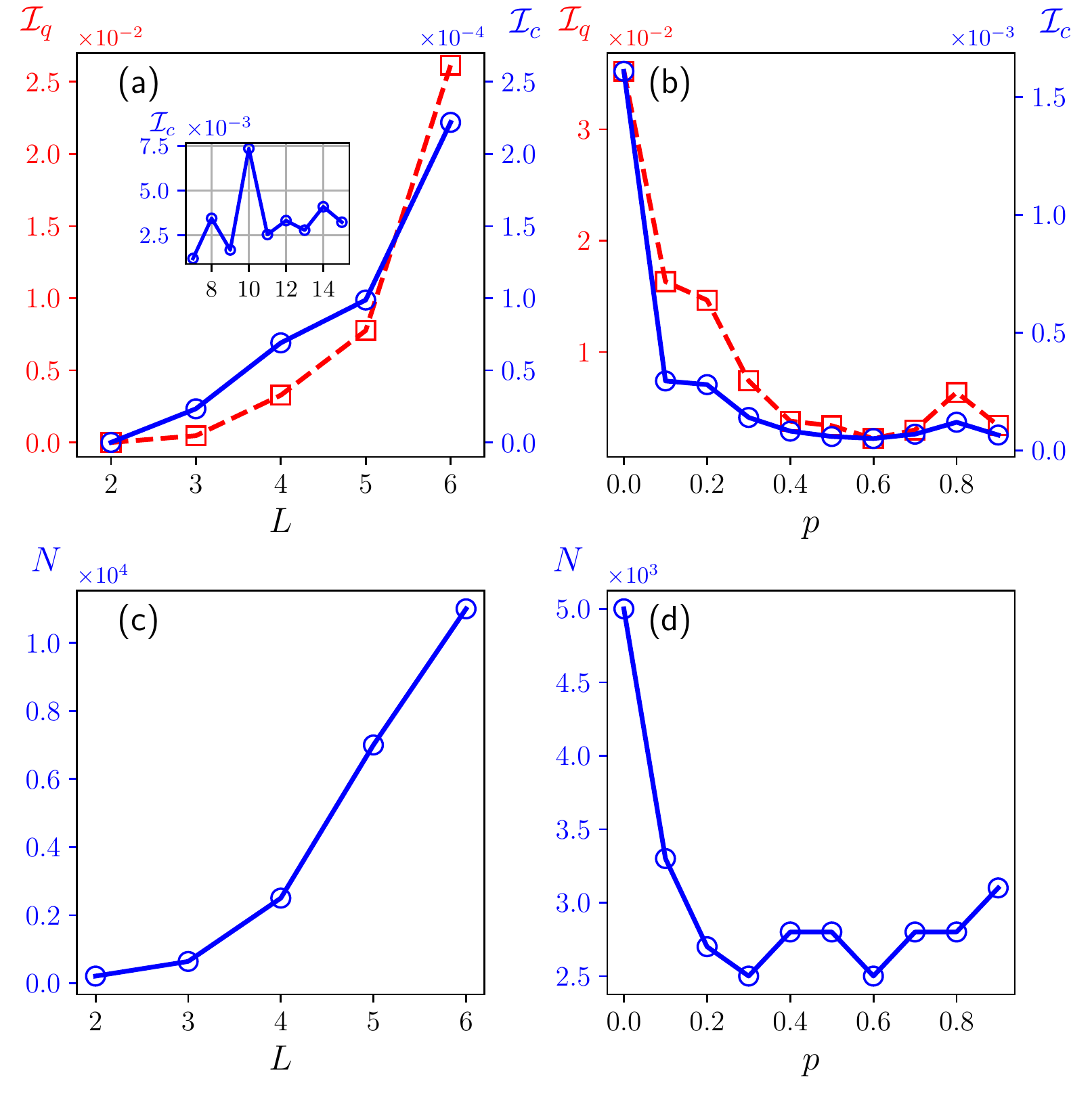}
\caption{(a) $\infidelity_q$ (red dashed line with squares) and $\infidelity_c$ (blue solid line with circles) as functions of system size $L$ for fixed $p=0.6$. The inset shows $\infidelity_c$ as a function of system size for larger system size ($\infidelity_q$ for $L>6$ is not shown since it is too expensive to compute). (b) $\infidelity_q$ and $\infidelity_c$ as functions of depolarized noise intensity $p$ for fixed $L=4$. (c) The minimum number of required training data $N$ as a function of $L$ such that $\infidelity_c \leq 1\%$, with $p=0.6$. (d) The minimum number of required training data $N$ as a function of $p$ such that $\infidelity_c \leq 1\%$, with $L=4$. All simulations are done with a bond dimension $D=10$.  }
\label{fig:fig1}
\end{figure}

\section{results}\label{sec:results}
We demonstrate our algorithm by reconstructing the density matrix corresponding to the ground state of the XXZ chain subjecting to depolarizing noise. The Hamiltonian of the XXZ chain can be written as
\begin{equation}\label{eq:ham}
    \hat{H} = \sum_{l=1}^{L-1}J(\hat{\sigma}^x_l\hat{\sigma}^x_{l+1} + \hat{\sigma}^y_l\hat{\sigma}^y_{l+1} + \gamma\hat{\sigma}^z_l\hat{\sigma}^z_{l+1}) + h\sum_{l=1}^{L}\hat{\sigma}^z, 
\end{equation}
where $L$ is the number of the spins, $J$ is the tunneling strength which we fix to $1$, $h$ is the magnetization strength, and $\gamma$ is the interaction strength. We choose $h=1$ to break the degeneracy of the ground state due to the spin flip symmetry. The depolarizing noise is described by the CPTP map:
\begin{equation}\label{eq:noise}
    \rhoop \rightarrow \mathcal{E} (\rhoop) = \frac{p\Iop}{d} + (1 - p)\rhoop,
\end{equation}
with $d=2^L$ as the dimension of the Hilbert space, $\rhoop$ the density matrix corresponding to the exact ground state, and $p$ the strength of the noise. We note that for $p=0$, namely for pure states, there already exists efficient MPS-based tomography algorithm which directly use MPS as the ansatz to represent an unknown pure state~\cite{cramer_efficient_2010}. 

\begin{figure}[htp]
\centering
\includegraphics[width=0.95\columnwidth]{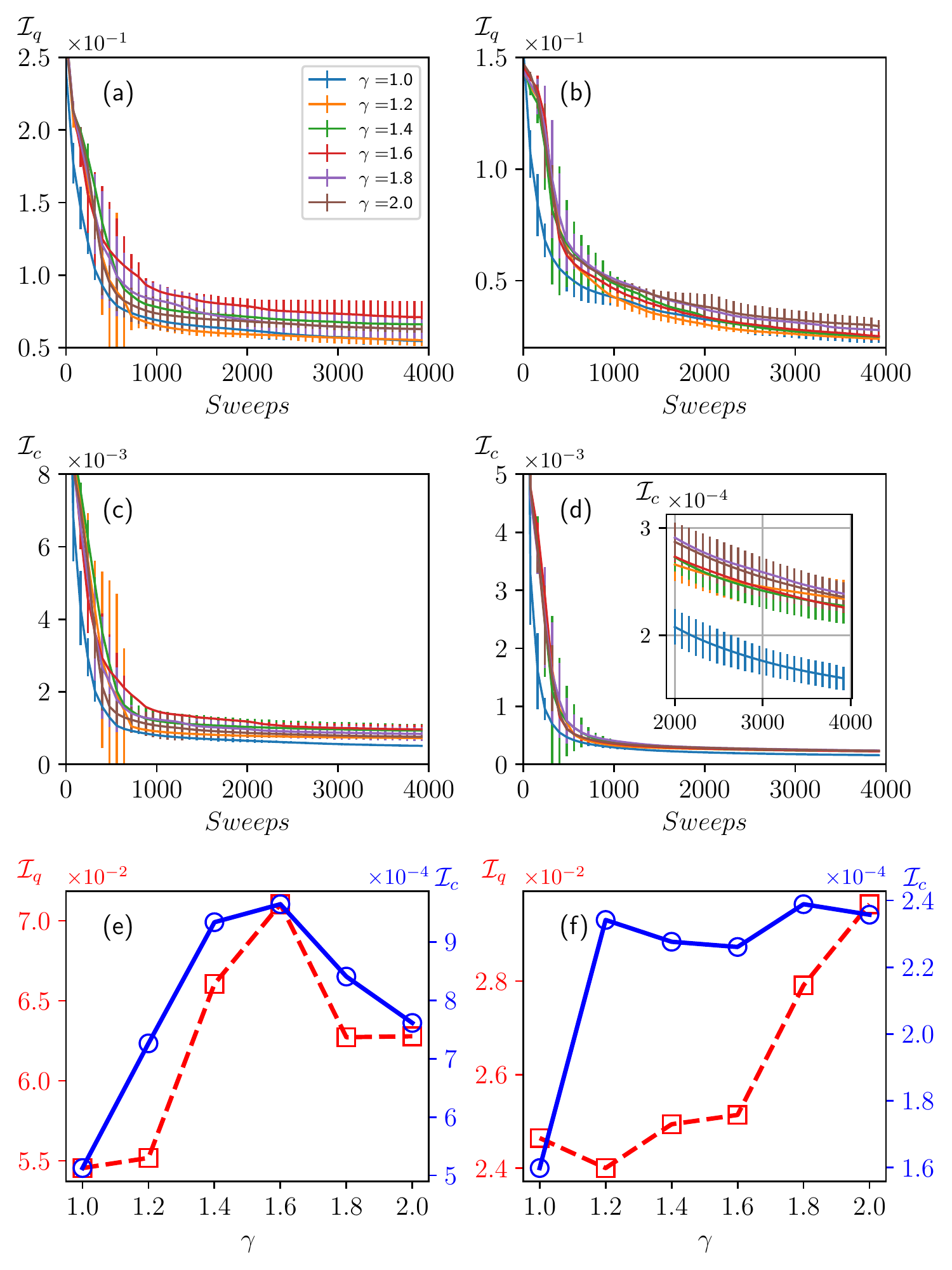}

\caption{\label{fig:fideity} (a, c) $\infidelity_q$ (a) and $\infidelity_c$ (c) as functions of the number of sweeps for different $\gamma$s with $p=0.4$. (b, d) $\infidelity_q$ (b) and $\infidelity_c$ (d) as functions of the number of sweeps for different $\gamma$s with $p=0.6$. The inset in (d) shows the tail of the convergence of $\infidelity_c$. In (a, b, c, d) we have chosen $5$ best results according to their loss values out of $100$ trials and plotted the mean values of them (The standard deviations are shown as error bars). (e) The left and right axis show the final $\infidelity_q$ and $\infidelity_c$ as functions of $\gamma$ with $p=0.4$. (f) The left and right axis show the final $\infidelity_q$ and $\infidelity_c$ as functions of $\gamma$ with $p=0.6$. Here we have chosen $L=6$ and used $N_{\train}=30000000$.
}
\label{fig:fig2}
\end{figure}

Similar to~\cite{carrasquilla_reconstructing_2019}, we use both the quantum fidelity and the classical fidelity to measure the learning accuracy. Specifically, the quantum fidelity is defined as
\begin{align}\label{eq:fq}
\fidelity_q = \trace^2\left(\sqrt{\sqrt{\rhoop_1}\rhoop_2 \sqrt{\rhoop_1}}\right),
\end{align}
for two density matrices $\rhoop_1$ and $\rhoop_2$, and the classical fidelity is defined as 
\begin{align}\label{eq:fc}
\fidelity_c = \mathbb{E}_{\bm{a}\sim P_i}[\sqrt{P_m(\bm{a})/P_i(\bm{a})}],
\end{align}
where $P_m(\bm{a})$ is the measured probability distribution, and $P_i(\bm{a})$ is the ideal probability distribution. We also define the quantum and the classical infidelities as $\infidelity_q = 1 - \fidelity_q $ and $\infidelity_c = 1 - \fidelity_c$ respectively. In our numerical simulations, we have generated two independent synthetic datasets for each parameter setting we have considered, each with $30,000,000$ samples. One dataset is used for training and the other is used for testing. For the quantum fidelity, we directly compute $\fidelity_q$ between the reconstructed density matrix and the target density matrix. For the classical fidelity, we use a test dataset to evaluate Eq.~(\ref{eq:fc}).

\begin{figure}[htp]
    \centering
    \includegraphics[width=0.9\columnwidth]{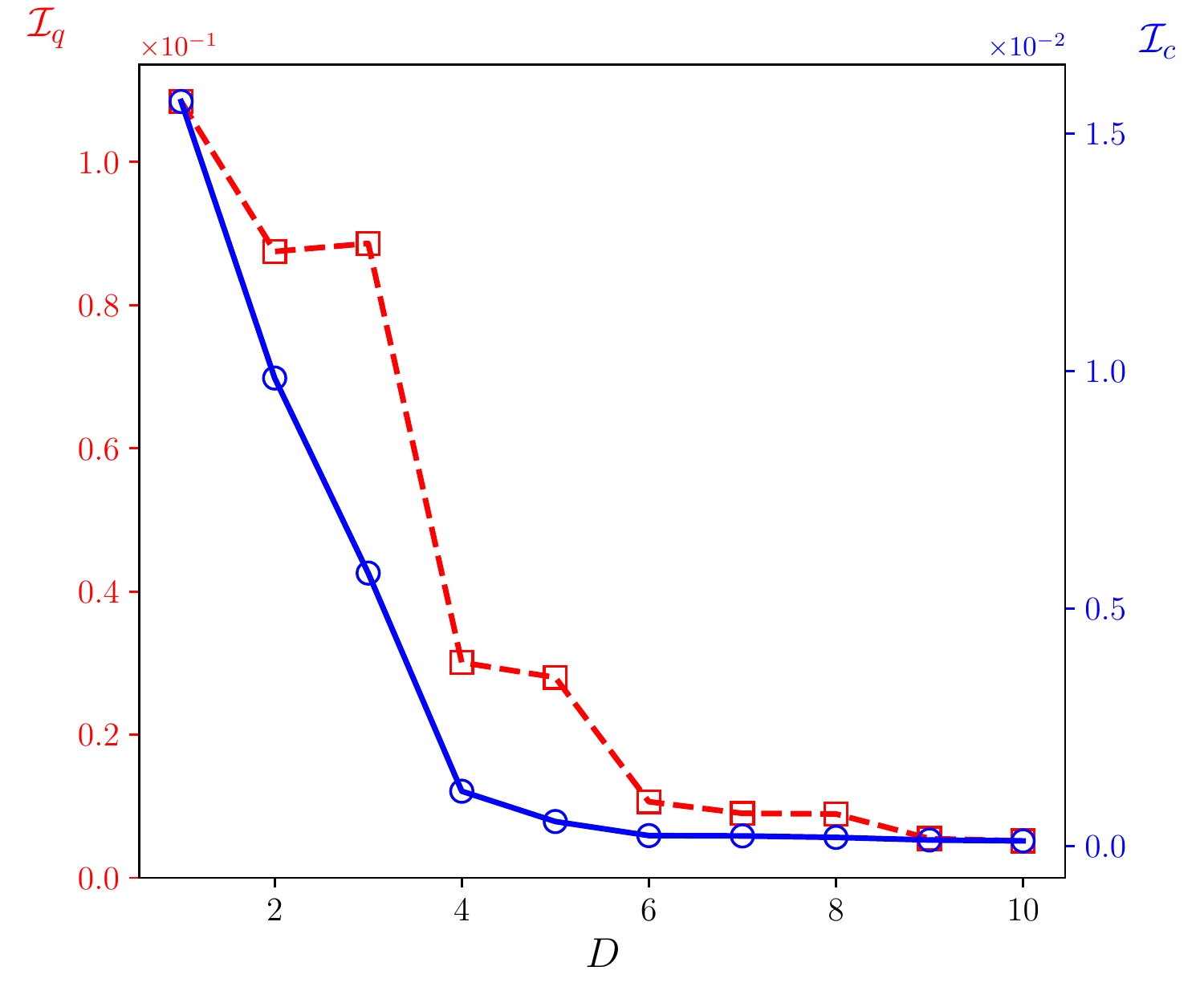}

\caption{$\infidelity_q$ (red dashed line with squares) and $\infidelity_c$ (blue solid line with circles) as functions of the bond dimension $D$. Here we have used $L=4$, $p=0.6$ and $N_{\train}=30000000$. }
\label{fig:fig3}
\end{figure}
We first study the final reconstruction quality as a function of the system size $L$ and the noise strength $p$. We show $\infidelity_q$ and $\infidelity_c$ as functions of $L$ in Fig.~\ref{fig:fig1}(a) and of $p$ in Fig.~\ref{fig:fig1}(b) respectively. We can see that the final fidelity (both the quantum and the classical) decreases as $L$ increases and increases as $p$ increases. We can also see that it is much easier for a near-perfect reconstruction of the probability distribution than the construction of the underlying quantum state, as in the numerical simulations $\infidelity_c$ is always at least one order of magnitude smaller than $\infidelity_q$. In Fig.(\ref{fig:fig1})(c, d) we show the minimum number $N$ of required training data for $\infidelity_c \leq 1\%$ as a function of $L$ and $p$ respectively. We can see that $N$ increases as $L$ increases and decreases as $p$ increases, as shown in  Fig.~\ref{fig:fig1}(a, b). 

Then we fix $L=6$ and investigate the variations of $\infidelity_q$ and $\infidelity_c$ as functions of the number of sweeps, which is shown in Fig.~\ref{fig:fig2}. In Fig.~\ref{fig:fig2}(a, c), we show $\infidelity_q$ and $\infidelity_c$ as functions of the number of sweeps when $p=0.4$ for different values of $\gamma$, while in Fig.~\ref{fig:fig2}(b, d) we show $\infidelity_q$ and $\infidelity_c$ as functions of the number of sweeps when $p=0.6$ for different values of $\gamma$. We can see that for both noise strengths, $\infidelity_c$ converges in about $1000$ sweeps and $\infidelity_q$ does not fully converge after $4000$ sweeps. The final values of $\infidelity_q$ and $\infidelity_c$ after $4000$ sweeps are also shown in Fig.~\ref{fig:fig2}(e) for $p=0.4$ and in Fig.~\ref{fig:fig2}(f) for $p=0.6$. We can see that in both cases $\infidelity_c$ is about two orders of magnitude smaller than the corresponding $\infidelity_q$ and that $\infidelity_q$ and $\infidelity_c$ for $p=0.6$ is smaller than those for $p=0.4$. %
\begin{figure}[htp]
    \centering
    \includegraphics[width=0.9\columnwidth]{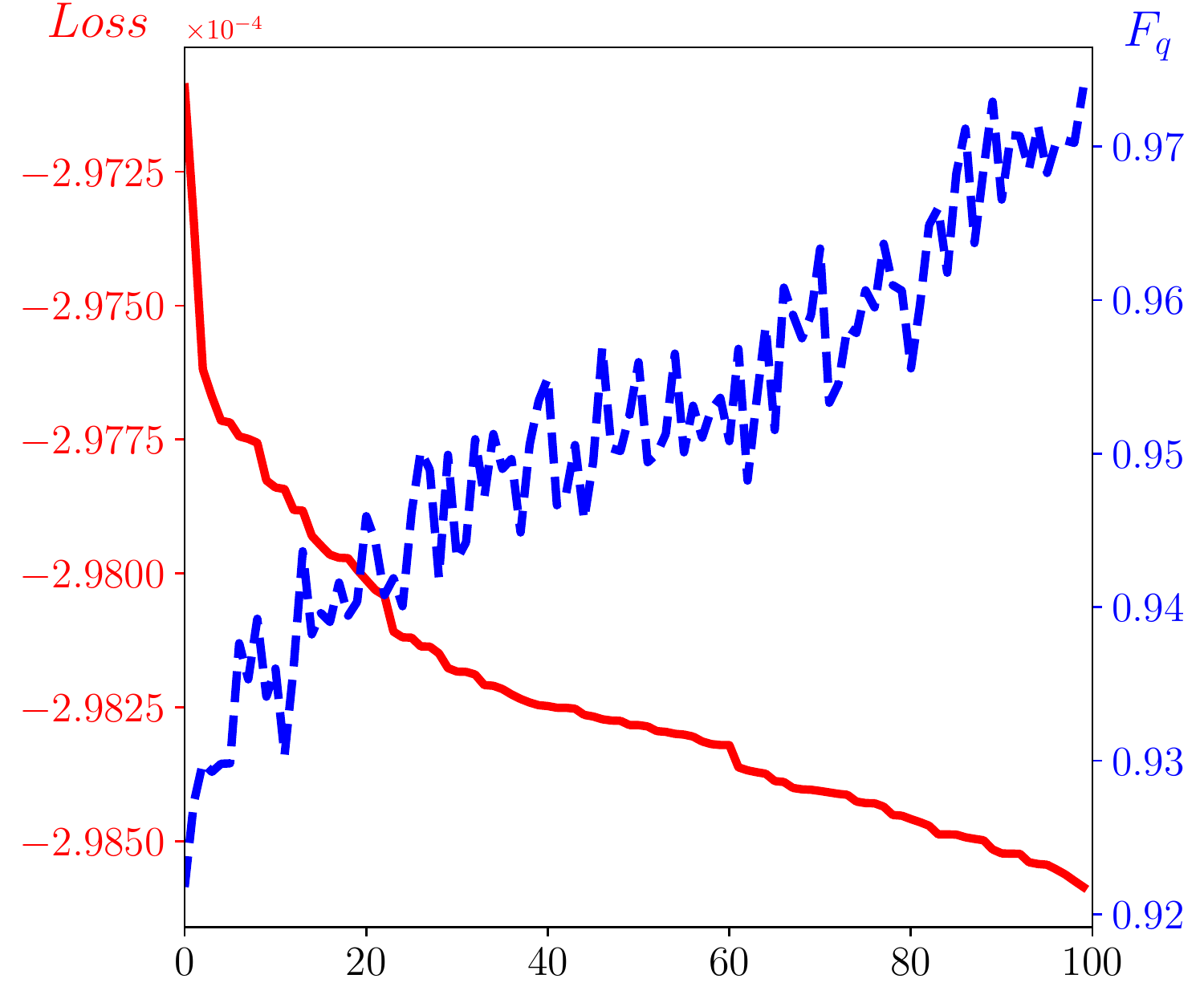}    
    \caption{\label{fig:loss} The $x$ axis denotes different labels of the $100$ numerical experiments, labelled $1$ to $100$, sorted by their final loss values from large to small. In our simulation, we have used $P^2(\bm{a}) - 2P(\bm{a})P_s(\bm{a})$ as the loss value, which simply shifts the original loss value in Eq.~(\ref{equ:fpa}) by a constant. Here the results are taken from the reconstruction of $P(\bm{a})$ for $L=6$, $\gamma=2.0$ and $p=0.6$. }  
    \label{fig:fig4}
    \end{figure}

Next we explore the learning accuracy as a function of the bond dimension $D$ used in our ansatz. The bond dimension of the MPS corresponds to the number of singular values retained in the process of restoring the original density matrix after matrix SVD. In our case the ground state can be represented as an MPS with a certain bond dimension $D_0$. For a perfect training, we need to have $D \geq D_0$ since otherwise our ansatz is not expressive enough to represent the target quantum state. If $D$ is too large and we do not have enough training data, we might have the problem of overfitting which would also result in bad learning accuracy. The dependence of the quantum and the classical infidelities as functions of $D$ are shown in Fig.~\ref{fig:fig3}, where we can see that both $\infidelity_q$ and $\infidelity_c$ decrease as $D$ increases until saturation.

Due to the variational feature of our algorithm similar to DMRG, it could be trapped in local minima (also because the initial MPS $P(\bm{a})$ is randomly initialized) [\onlinecite{LiuDeng2021}]. Therefore in our numerical results, the same reconstruction algorithm is run for many trials with random initialization of $P(\bm{a})$, and the one with the lowest loss value is chosen as the final result. Ideally one would likely to directly choose the trials with the highest fidelity. However in real experiment the target state is unknown and it is not possible to compute the fidelities. As a result it is important that the trials with lower loss values will correspond to those with higher fidelities. Such correspondence between loss values and fidelities is shown in Fig.~\ref{fig:loss}, where we have repeated the reconstruction algorithm for $100$ times. We can see that indeed the loss value has the desired correspondence with the fidelity.

\section{Conclusion}\label{sec:conclusion}
We have presented an algorithm based on the non-negative matrix product state for quantum state tomography. Given a number of experimental measurement outcomes, our algorithm iteratively finds the optimal non-negative MPS representation which best approximates the probability distribution of these outcomes. Applying simple local transformations, the reconstructed non-negative MPS can be converted into a density matrix for the unknown quantum state. This is in comparison with the QST methods based on neural network states, for which one generally can not directly write down the quantum state but only has indirect access to it via sampling from the trained neural networks. As applications, the effectiveness of our algorithm is demonstrated to reconstruct the ground state of the XXZ chain with depolarizing noise.

\begin{acknowledgments}
CG acknowledges support from National Natural Science Foundation of China under Grants No. 11805279. DH and XW gratefully acknowledge the grant from the National Key R\&D Program of China, Grant No. 2018YFA0306703.
\end{acknowledgments}


\bibliography{tomo_MPS}

\end{document}